\documentclass[aps,showpacs,twocolumn,longbibliography,nofootinbib,superscriptaddress]{revtex4-1}
\usepackage{graphicx} 
\usepackage{amssymb}
\usepackage{float}
\usepackage[T1]{fontenc}
\begin{document}
\title{Enhanced spontaneous down-conversion in a nonlinear crystal \\ embedded with  plasmonic-quantum emitter hybrid structures}

\author{Zafer Artvin}
\affiliation{Institute  of  Nuclear  Sciences, Hacettepe University, 06800 Ankara, Turkey}
\affiliation{Department of Nanotechnology and Nanomedicine, Hacettepe University, Ankara, Turkey}
\affiliation{Central Laboratory, Middle East Technical University, 06800 Ankara, Turkey}
\author{Mehmet Emre Tasgin}
\affiliation{Institute  of  Nuclear  Sciences, Hacettepe University, 06800 Ankara, Turkey}
\author{Alpan Bek}
\affiliation{Department of Physics, Middle East Technical University, 06800 Ankara, Turkey}
\author{Mehmet Gunay}
\affiliation{Department of Nanoscience and Nanotechnology, Faculty of Arts and Science, Mehmet Akif Ersoy University, 15030 Burdur, Turkey}
\begin{abstract}
Control of nonlinear response of nanostructures via path interference effects, i.e. Fano resonances, has been studied extensively. In such materials, a frequency conversion process which takes place near a hot spot has been considered. Here, we study a different case. The frequency conversion process takes place along the body of a nonlinear crystal.  Metal nanoparticle-quantum emitter dimers control the down-conversion process, taking place throughout the crystal body, via introducing interfering conversion paths. Dimers behave as interaction centers. We show that a 2 order of magnitude enhancement is possible, beyond an enhancement due to localization effects. This factor multiplies the enhancement taking place due to the field localization.
\end{abstract}
\maketitle

\section{INTRODUCTION}\label{sec:level1}

Spontaneous down conversion~(SDC), a second-order nonlinear process, splits an incident beam into two subfrequencies labeled as signal and idler~\cite{BOYD}. Nonlinear materials,  like barium borate~\cite{IEEE14}, LiNbO3~\citep{libo3}, are employed in generating these photon pairs, due to their large nonlinearities. Down-conversion process can entangle two down-converted beams/photons in various degrees of freedom: such as continuous-variable, polarization, space, time and orbital angular momentum~\cite{LightScienceApplications9}. Entangled beam/photon pairs are essential for many fundamental quantum optics experiments~\cite{PRL10} as well as a key resource in quantum communication including cryptography~\cite{PRL11}, quantum computation~\cite{PRL12} and quantum information~\cite{euler2011spontaneous}. SDC can also be used in solar cell applications by means of reducing the energy losses with converting high energy photons into two or more lower energy photons~\cite{trupke2002improving,JAP17}.


Despite such important implementations in quantum technologies~\cite{NaturePhot_2019_IntegratedQuantumTechnologies,NaturePhot_2009_QuantumTechnologies,Nanophot_2017_PlasmonicsQuantumTechnologies}, limited efficiencies of materials in parametric-down conversion process, i.e. the conversion rate, still poses a problem in technological applications~\cite{PRL10}. Some recent works~\cite{SR18, OE19} reported on improvement of the conversion efficiency via changing the geometry/structure of the nonlinear crystals. Some experimental works employ interaction of the crystal with quantum dots~(QDs), as quantum emitters~(QEs). QDs, embedded into photonic crystal structures~\cite{qd} and into a nonlinear down-converting crystal, resonant with the down-converted mode~\cite{cavitysdc,bilgelt}, are shown to enhance the parametric-down conversion process. Some more recent works utilize the plasmonic localization effect of metal nanoparticles. For instance, recent theoretical studies~\cite{Plasmon_th1,Plasmon_th2} envision a 5 orders of magnitude enhancement in SDC efficiency when plasmonic nano-structures are embedded into nonlinear crystals. These results make one draw a conclusion that: plasmonic nanoparticles, localizing (concentrating) the field into hot spots (in the crystal medium), can help nonlinearity~\cite{plasnap,gianten,cuzn,libo3}. Because a nonlinear response increases with the second power of the local intensity. 
Experiments on the integration of plasmonic nanostructures into nonlinear devices, such as fiber glasses~\cite{fiber2} and photonic crystals~\cite{pc2,pc3} also report enhanced nonlinear response, e.g. on the Raman conversion process. A second harmonic generating cavity, embedded with plasmonic nanostructures, is also experimentally demonstrated to yield enhanced sum-frequency generation~\cite{
SHGcrystal_MNPs_Nanoscale2018,MNPCrystalNonlineEnhance_ACSOmega_2017,mukherjee2011one,bigot2011linear}.

The field-localizing feature of the metal nanostructures not only enhances the nonlinear processes, but it also leads to enhanced light-matter interaction at the hot spots. Quantum emitters~(QEs), placed at the hot spots, couple to the concentrated near-field (polarization) of the plasmon excitation, which is several orders of magnitude stronger compared to their direct coupling to the incident light. Strong interaction between a QE and a metal nano particle (MNP) makes path interference effects visible: Fano resonances~\cite{ridolfo2010quantum}, the plasmon-analog of electromagnetically induced transparency~(EIT)~\cite{scully1999quantum}. Similar to EIT-like behaviors in atomic clouds, Fano resonances can be used to control the refractive-index~\cite{panahpour2019refraction,gunay2019tunable} and nonlinear conversion processes~\cite{butet2014fano,turkpence2014engineering,selen}. The origin of these interference effects, e.g. enhancement and suppression of both  linear and nonlinear response, can be demonstrated with a basic analytical model, where cancellations in the denominator of a converted amplitude result in enhancement of the processes. Besides providing such control methods on the steady-state amplitudes,  Fano resonances can also increase the lifetime of plasmon oscillations~\cite{FR1,FR2,bilgelt}, leading a further accumulation of the field intensity at the hot spots, i.e. dark-hot resonances \cite{darkhot,zhang2013coherent,zhang2014coherent}.

Such plasmonic path interference effects, where the nonlinear conversion process takes place on a local region (e.g. a nanoparticle or a molecule), have already been studied extensively. In these setups both the generation of the nonlinear (up or down conversion) field and the interaction of the quantum emitter with such a nano-converter take place at the hot spot. In systems, where nanoparticles are embedded into nonlinear crystals~\cite{plasnap,gianten,cuzn,libo3}, however, nonlinear process takes place all over the crystal body, i.e. not merely on the nanoparticle hot spot. 

In this work, we study the nonlinear response of a down-converting crystal into which MNP-QE dimers are embedded, see Fig. 1.a.~MNPs behave as interaction centers. They make the unlocalized down-converted field concentrate into the hot spots where a stronger interaction with a quantum emitter takes place, e.g., compared to QE-embedded crystals~\cite{plasnap,gianten,cuzn,libo3}. We show that such a setup can enhance the down-converted field 2 orders of magnitude. It should be emphasized that, this enhancement factor comes as a further multiplication factor on top of the MNP hot spot enhancement.

 multiplies the enhancement appearing due to the MNP-localization of the pumped cavity mode~\cite{SHGcrystal_MNPs_Nanoscale2018,MNPCrystalNonlineEnhance_ACSOmega_2017,mukherjee2011one,bigot2011linear}.

\begin{figure}
\includegraphics[width=8cm, height=6cm]{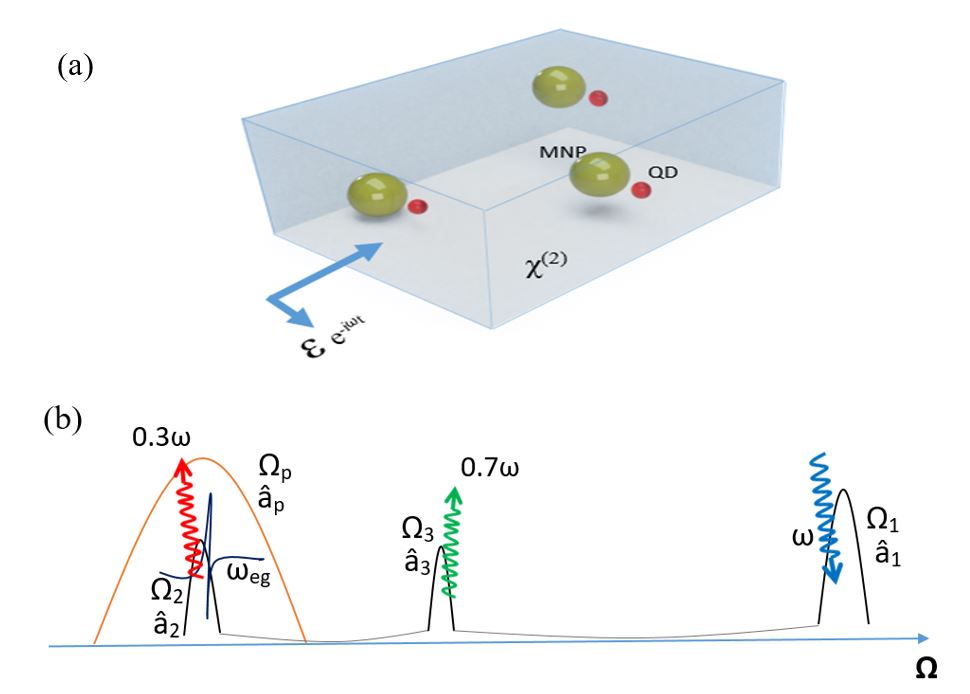}
\label{Sketch}
\caption{(a) A nonlinear crystal performs down conversion process. Metal nanoparticle~(MNP)-quantum dot~(QD) dimers are embedded into the crystal to increase the conversion efficiency. The MNPs behave as interaction centers. They localize the already generated $\omega_2=0.3\omega$ field and introduce path interference effects. This controls the amplitude $\alpha_2$ of the $\omega_2=0.3\omega$ field. (b) Cavity modes of the nonlinear crystal, $\Omega_{1,2,3}$, which supports the pumped, $\omega_1=\omega$, and the down converted $\omega_2=0.3\omega$, $\omega_3=0.7\omega$ modes. Down-converted $\omega_2 = 0.3\omega$ photons interact with the MNP-QE hybrid structures. $\Omega_p$ is the resonance of MNP plasmons. Crystal modes irrelevant with the down-conversion process have not been depicted. Polarization of the pump$~ E~ e^{-i\omega t}$ field, determining also the polarization of the down-converted field, is chosen along the dimer axis.}
\end{figure} 

The laser pump excites the cavity mode $\Omega_1$ of the nonlinear crystal, i.e. oscillating as $\propto e^{-i\omega t}$. The $\omega$ oscillations in the $\Omega_1$ cavity mode are down-converted into $\omega_2=0.3\omega$ and $\omega_3=0.7\omega$ oscillations in the cavity modes $\Omega_2$ and $\Omega_3$, see Fig.~1.b. We make the cavity field $\omega_2=0.3\omega$ interact with the MNP's plasmon mode ( $\Omega_p$) and introduce a path interference in this converted mode. We examine the behaviour of the down-converted field for different coupling parameters, i.e. between the $\omega_2=0.3\omega$ oscillations in the $\Omega_2$ cavity mode and the MNP, $g$, and the one between the MNP and the QE, $f_c$, see Fig.~3. We define the enhancement factors by comparing the $\omega_2=0.3 \omega$ down-converted intensities when the MNP-QE is absent with the case when the MNP-QE is present.

\begin{figure}[h]
 \centering
\includegraphics[width=7cm, height=3cm]{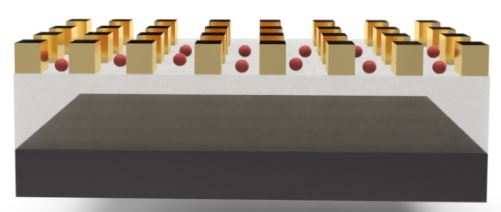}
\caption{MNP-QE dimers can be coupled to the down-converted $\omega_2 = 0.3\omega$ field via evanescent waves. Dimers can be controllably positioned on the mode lobe maxima. Controlled positioning of dimers~\cite{bek2008fluorescence} can also work in the favor of cooperative operation of path interferences from many dimers. Such a setup also enables the tuning of the quantum dot level spacing ($\omega_{eg}$) via an applied voltage.} 
\label{fig2}
\end{figure} 

We introduce a basic analytical model including the damping rates of the three cavity modes, $\gamma_{1,2,3}$, and the one for the MNP plasmon mode $\gamma_p$. We use realistic values for the damping rates one can observe in typical experiments. This is similar for couplings $g$ and $f_c$. The setup, considered in Fig.~1a, can be obtained via ion implantation techniques, where accelerated MNPs and QDs are targeted into the nonlinear crystal orientation of dimers would be random by using this technique~\cite{SHGcrystal_MNPs_Nanoscale2018,MNPCrystalNonlineEnhance_ACSOmega_2017,mukherjee2011one,bigot2011linear}. 

An alternative and probably more feasible experimental scheme can be seen in Fig. 2. Interaction of the converted field $\omega_2=0.3\omega$ with MNP-QE dimers can also be achieved via evanescent fields. Lithography techniques~\cite{altissimo2010beam} can be used to fabricate periodic metal nanostructures on the nonlinear crystal surface and QDs can be deposited between these metal nanostructures (see Fig. 2). When the dimers are positioned on the mode lobe maxima of the converted cavity mode, evanescent radiation couples to the dimers, see Fig. 2. In such a case, the level-spacing of the quantum dots can also be tuned continuously via an applied voltage. Size, shape, type of material and the quantity of embedded nanoparticles affect the nonlinear properties of the crystal. 
 
While here we consider QEs for longer lifetime particles, coupled to the MNPs, they can also be replaced with metal nanostructures supporting long-life dark-modes. Actually, metal nanostructures, or even a single MNP structure, can support both bright and dark modes where the two modes interact with each other and result in Fano resonances~\cite{Remo2014dark,bilgelt}.

We, in advance, note that the analytical model, we base our results on, does not take the retardation effects into account. 3D FDTD solutions of Maxwell equations show that retardation effects do not destroy the appearance of Fano resonances when the nonlinearity takes place on (or close to) a nanoparticle. The case we study here, however, considers a nonlinearity conversion which takes place all over the nonlinear crystal, i.e. not only on the hot spot. Thus, while we are aware that retardation effects have to be revisited~\footnote{\label{fn:3Dsimulations} Second harmonic process is encoded in many FDTD programs through the exact form of the nonlinear 3D Maxwell equations. The down-conversion process (also, e.g., Raman conversion process), however, is not treated with exact solutions of nonlinear Maxwell equations, but their enhancement is predicted from the localization factors~\cite{plasnap,cuzn,gianten}.}, our aim in this paper is solely to study the path interference effects on the parametric down conversion process.

Actually, in a new study~\cite{gunay2019fano}, we just recently managed to show that retardation effects do not wash out the path interference phenomenon. In Ref.~\cite{gunay2019fano}, we consider a setup, similar to the one here in Fig.~1, but inspect the second harmonic generation process. More explicitly, solutions of 3D nonlinear Maxwell equations show that an extra enhancement due to path interference multiplies the enhancement due to localization (of MNPs). In the present work, we concentrate on the enhancement factors taking place due to the path interference~\footnote{3D simulations of Ref.~\cite{gunay2019fano} shows that such effects are not washed out by retardation effects.}. Our basic model cannot take into account the enhancements due to localization; thus, presents the ones due to path interference only.

The paper is organized as follows. In Sec.~\ref{sec:model}, we describe our model and the mechanism for the spontaneous down conversion process in a nonlinear crystal medium. We derive the hamiltonian and the equations of motion for the plasmon and down-converted modes. In the Sec.~\ref{sec:results}, we present the enhancement factors for various interaction strengths ($f_c, g$) and QD level-spacing $\omega_{eg}$. We also examine the plasmon occupation of the MNPs. Sec.~\ref{sec:conclusion} contains our summary and discussions.

\section{ The model} \label{sec:model}

In this section, we first derive an effective Hamiltonian for a down-converting crystal loaded with MNP-QE hybrid structures. We then obtain the equations of motion for this system and time evolve the equations to obtain the steady-states of the plasmon and down-converted fields.

The setup we consider is depicted in Fig. 1. An incident light of frequency $\omega$ interacts with the nonlinear crystal and excites the $\hat{a}_1$-mode of the crystal. Down conversion process, $0.3
\omega$ and $0.7\omega$, takes place into two other cavity modes $\hat{a}_2$ and $\hat{a}_3$, respectively. The resonances of the cavity modes $\hat{a}_{1,2,3}$ are $\Omega_{1,2,3}$. The low-energy down-converted field $0.3\omega$, in the $\hat{a}_2$ mode, interacts with the embedded MNPs and excites surface plasmons ($\hat{a}_p$ mode). The interaction strength between the $\hat{a}_2$ mode and the MNP is $g$. A QE of level spacing $\omega_{eg}$ is placed at the hot spot of the MNP. The level-spacing of QE is chosen close to the plasmon field oscillations ($ \omega_{eg}  \sim \Omega_p$). Thus, QE interacts only with the $\hat{a}_2$ mode. Here, due to the enhanced hot spot intensity, there appears a strong interaction between the MNP and the QE ($f_c$). The interaction of the QE with the $\hat{a}_2$ mode, the $0.3~\omega$ down-converted field, takes place over the intense MNP hot spot, thus, the direct interaction between the $\hat{a}_2$ mode with the QE is negligible~\cite{turkpence2014engineering}, see Fig. 3. 

The interaction hamiltonian for the down conversion process, taking place in the nonlinear crystal medium, can be written as
 \begin{eqnarray}\label{e:barwg}
 \hat{\cal{H}}_{\rm dc}&=&\int d^3r \chi^{(2)}  \Bigl[\hat{a}_3^\dagger\hat{a}_2 ^\dagger\ \hat{a}_1 E_1(\textbf{r}) E_2 ^\ast(\textbf{r}) E_3^\ast(\textbf{r})   \nonumber \\
&+& E_1^\ast(\textbf{r}) E_2(\textbf{r}) E_3(\textbf{r})  \hat{a}_1^\dagger \hat{a}_2 \hat{a}_3  \Bigr ],
 \end{eqnarray}
where $  E_1^\ast(\textbf{r})$ ($ E_2(\textbf{r}) $ $E_3(\textbf{r}) $) are the positive (negative) frequency parts of the electric fields. $ \hat{a}_j^{\dagger} $ ($ \hat{a}_j $) is the creation (annihilation) operator of the j-th mode. $ \chi_{2}= \int d^3r \chi^{(2)} E_1^\ast(\textbf{r}) E_2(\textbf{r})  E_3(\textbf{r})$ is an overlap integral, which determines the strength of the down conversion process. Here, one can consider $\chi_{2} $ as a 3D step function which is zero outside the crystal body.

The total hamiltonian of the system can be written as the sum of the energies of the crystal oscillations, i.e. the pumped~($ \Omega_1 $) and the down converted~($\Omega_2$ and $\Omega_3$) crystal fields, energy of the plasmon oscillations~($ \Omega_p $), energy of the two-level QE~($ \omega_{eg} $) and the energy transferred by
the laser source~($  \varepsilon e^{-i\omega t}  $). 
\begin{eqnarray}
\hat{\cal{H}}_0&=& \sum_{i=1}^{3} \hbar\Omega_{i} \hat{a}_{i}^\dagger \hat{a}_i   +\hbar \Omega_p \hat{a}_{p}^\dagger \hat{a}_p ,
+\hbar \omega_{eg} |e \rangle \langle e| \nonumber \\
\hat{\cal{H}}_{L}&=&i\hbar (\varepsilon  \hat{a}_1^\dagger e^{-i\omega t} -\textit{H.c.}).
\label{Ham0}
\end{eqnarray} 
Here, $|g \rangle$~($|e \rangle $) is the ground (excited) state of the QE. $ \hat{\cal H}_{\rm int}$ contains  the interactions between the down-converted field and the plasmon mode, $g$, and the interaction between the MNP (plasmon mode) and the QE, $f_c $. The parameters for the coupling strengths depend on the relative positions of the particles and the spatial overlaps of the cavity and plasmon modes and the overlap of the QE with the hot spot. Thus, the down conversion process $\hat{H}_{\rm dc}$ and the interaction hamiltonian can be written as
\begin{eqnarray}
\hat{\cal{H}}_{\rm dc} &=& \hbar \chi_{2} (\hat{a}_1^\dagger \hat{a}_2 \hat{a}_3+ \hat{a}_2^\dagger \hat{a}_3^\dagger \hat{a}_1 ),
\\  
\hat{\cal{H}}_{\rm int}&=&\hbar f_c (\hat{a}_p^\dagger |g\rangle \langle e|+ |e\rangle \langle g| \hat{a}_p )+\hbar g(\hat{a}_p^\dagger \hat{a}_2+\hat{a}_2 ^\dagger\hat{a}_p). 
\end{eqnarray}

Dynamics of the system can be obtained using the Heisenberg equations of motion (e.g. $i\hbar\dot{\hat{a}}_i=[\hat{a}_i,\hat{\cal H}] $). Since we are interested in the intensities only, but do not aim to calculate the correlations, we replace the operators $\hat{a}_i$ and $ \hat{\rho}_{ij}= |i\rangle \langle j|$ with complex (c-) numbers  ${\alpha}_i$ and $ {\rho}_{ij} $ respectively~\cite{premaratne2017theory}. The equations of motion can be obtained as
\begin{eqnarray}
\dot{{\alpha}_1}&=&-(i\Omega_1+\gamma_{1}) {\alpha}_1-i  \chi_{2} {\alpha}_2 {\alpha}_3 + \varepsilon e^{-iwt}\label{EOMa},\\
\dot{{\alpha}_2}&=&-(i\Omega_2+\gamma_{2}) {\alpha}_2-i  \chi_{2} {\alpha}_1 {\alpha}_3^{\ast} - i g {\alpha}_{p} \label{EOMb}, \\
\dot{{\alpha}_3}&=&-(i\Omega_3+\gamma_{3}) {\alpha}_3-i  \chi_{2} {\alpha}_1 {\alpha}_2^{\ast} \label{EOMc}, \\
\dot{{\alpha}}_p &=&  -(i \Omega_p+\gamma_{p}){\alpha}_p-i g {\alpha}_2-if_c{\rho}_{ge} \label{EOMd} ,\\
\dot{{\rho}}_{ge} &=&  -(i \omega_{eg}+\gamma_{eg}){\rho}_{ge}+i f_c{\alpha}_p({\rho}_{ee}-{{\rho}}_{gg}) \label{EOMe},\\
\dot{{{\rho}}}_{ee} &=& -\gamma_{ee} {{\rho}}_{ee}+i (f_c {\rho}_{ge} {\alpha}^\ast_p- \textit{c.c}),
\label{EOMf}
\end{eqnarray}
where $\gamma_{1,2,3}$, $ \gamma_{p} $ and $ \gamma_{ee} $  are the damping rates of  crystal modes, the plasmon mode and the quantum emitter, respectively. There exists also a constraint for the conservation of probability, i.e., $ {{\rho}}_{ee} +{{\rho}}_{gg} =1$. $ \gamma_{eg} = \gamma_{ee}/2 $ is the off-diagonal decay rate for a single QE. We carry out the time evolution of the differential equations, Eqs.~(\ref{EOMa})-(\ref{EOMf}), to obtain the steady state amplitudes.

\begin{figure}[h]
\centering
\includegraphics[width=8 cm, height=3 cm]{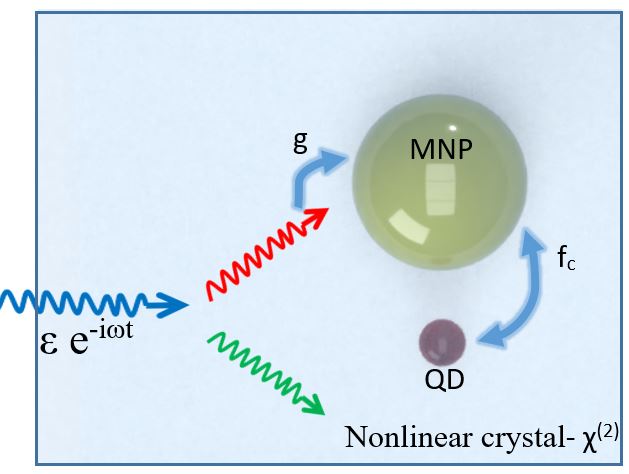}
\label{Sketch}
\caption{A sketch demonstrating the down-conversion and the interaction processes taking place in the nonlinear crystal. The input laser pumps the $\omega$ oscillations in the $\hat{a}_1$ (crystal) cavity mode. $\omega$ oscillations in the cavity is down-converted into $0.3\omega$ ($\Omega_2$ mode) and $0.7\omega$ ($\Omega_3$ mode) oscillations in the crystal. The $0.3\omega$ oscillations couple (strength $g$) with the MNP whose plasmon mode ($\Omega_p$) is around $0.3\omega$. A quantum dot~(QD), at the hot spot of the MNP, couples to the plasmon oscillations, at the frequency $0.3\omega$. Direct coupling of the QD to the $\Omega_2$ mode, $0.3\omega$ field, is small compared to its coupling to $0.3\omega$ oscillations over the hot spot.}
\end{figure}

\section{Results} \label{sec:results}
 The enhancement factor for the down-converted field intensity,
\begin{eqnarray}
{\rm  Enhancement\quad Factor}=\frac{|\alpha_2 (f_c\neq 0,g\neq 0)|^2}{|\alpha_2(f_c=0,g = 0)|^2} ,
\label{EF}
\end{eqnarray}
is defined as the ratio of the down converted intensities in the presence~($ f_c\neq 0,g\neq 0 $) and absence ($ f_c=0,g=0 $)  of the MNP-QE hybrid structure. We calculate the enhancement factors at the steady-state, from the time evolution Eqs. (\ref{EOMa}-\ref{EOMf}) and using Eq.(\ref{EF}). We investigate the effect of MNP-QE hybrid structure on the output signal of the down converting crystal. 

\begin{figure}
\begin{center} 
 \includegraphics[width=6.5cm, height=4.2cm]{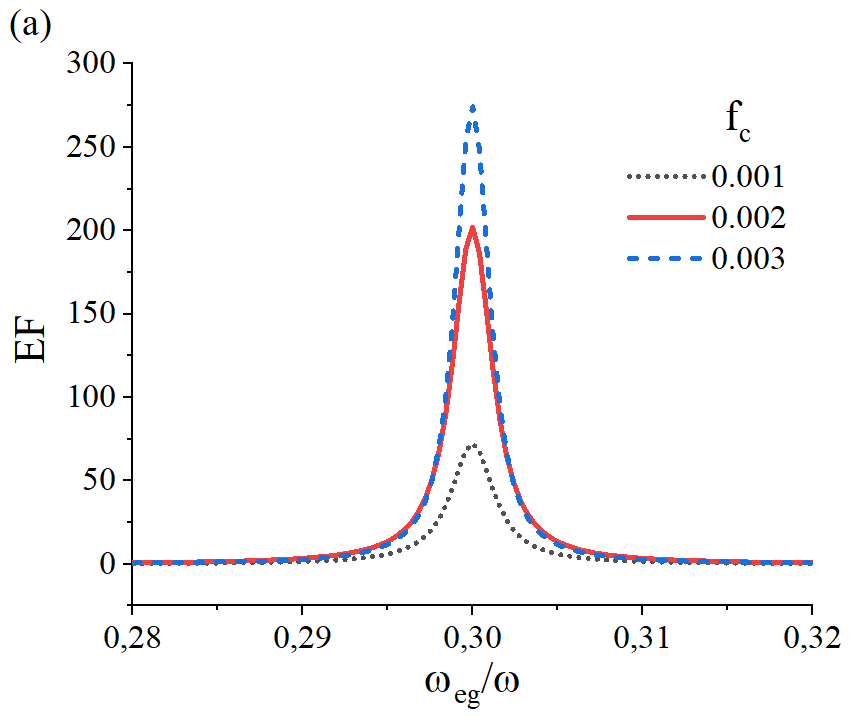}
\includegraphics[width=6.5cm, height=4.2cm]{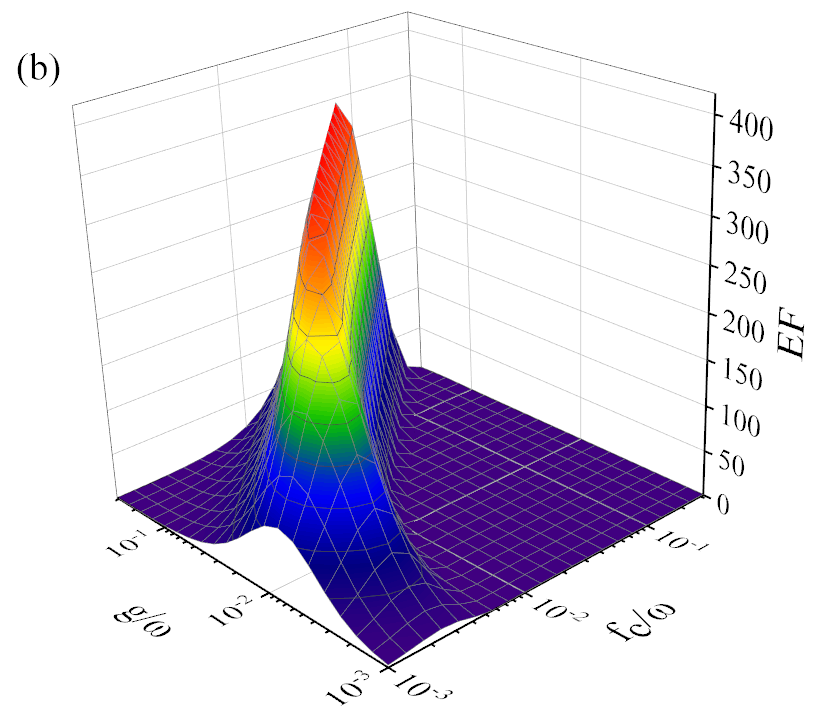}
\caption{Enhancement factor~(EF) of the down converted $0.3\omega$ signal intensity (a) versus QE's level-spacing $\omega_{eg}$ calculated for different MNP-QE interaction strengths  $f_c$. We set $g=0.01\omega$ for the coupling of the MNP to the $\alpha_2$ crystal field. $\alpha_2$ supports the $0.3\omega$ oscillations. (b) EF, at fixed $\omega_{eg}=0.3\omega$, calculated for different $f_c$ and $g$ couplings.  There appears 300 to 400 fold EFs, crudely, for the ratios $f_c/g\sim 1/4$. Other parameters we useare $\Omega_1=\omega$, $\Omega_2=0.3\omega$, $\Omega_3=0.7\omega$, $\Omega_p= 0.3\omega$, $\gamma_1=\gamma_2=\gamma_3=5\times10^{-4}\omega$, $\gamma_{p}=0.1~\omega$, $\gamma_{eg}=10^{-5}~\omega$ and $  \chi_2=2\times 10^{-9}~\omega$.}
\end{center}
\end{figure}  

We examine the down converted $\omega_2=0.3\omega$ field intensity, $|\alpha_2|^2$, for different interaction strengths $g$ and $f_c$. These parameters change with MNP sizes and the position of the quantum emitter~(QE). Additionally, in the alternative setup given in Fig.~\ref{fig2}, QD level spacing $\omega_{eg}$ can also be continuously tuned via an applied voltage. In Fig.~\ref{fig2}, MNPs and QDs are positioned on the surface of the crystal and they are coupled to the down converted field over the evanescent waves.

Fig. 4a shows that path interference phenomenon can achieve 2-orders of magnitude enhancement factors in the presence of the MNP-QD dimers, despite the fact that the damping rate of the MNP $\gamma_p$ (coupled to the $\hat{a}_2$ mode) is orders of magnitude larger than the one for the $\alpha_2$ crystal mode. We scale the frequencies with $\omega$, the frequency of the laser pumping the $\alpha_1$ crystal mode. We use $g=0.01\omega$, $\Omega_1=\omega$, $\Omega_2=0.3\omega$, $\Omega_3=0.7\omega$, $\Omega_p= 0.3\omega$, $\gamma_1=\gamma_2=\gamma_3=5\times10^{-4}\omega$, $\gamma_{p}=0.1\omega$  and $\gamma_{eg}=10^{-5}\omega$. We consider a small (arbitrary) second order nonlinear (overlap) susceptibility  $\chi_{2}=2\times 10^{-9}\omega$ whose actual value does not change the enhancement ratios.

Fig. 4b demonstrates how the interaction strengths between MNP and QE~($ f_c $) and between the MNP plasmon field and the down-converted $\alpha_2$ field~($g$) affect the enhancement factor for a fixed QE level spacing~($ \omega_{eg}=0.3\omega$). We observe that enhancement takes place at a certain ratio of $f_c/g$. subject to the condition $f_c<g$. We can crudely determine this ratio as $f_c/g\sim 1/4$, where enhancement of the $|\alpha_2|^2$ down converted intensity appears. As the $f_c$ value gets closer to the $g$ and  $f_c>g$, enhancement factor (EF) decreases.

\begin{figure}[h]
 \centering
\includegraphics[width=7cm, height=4.2cm]{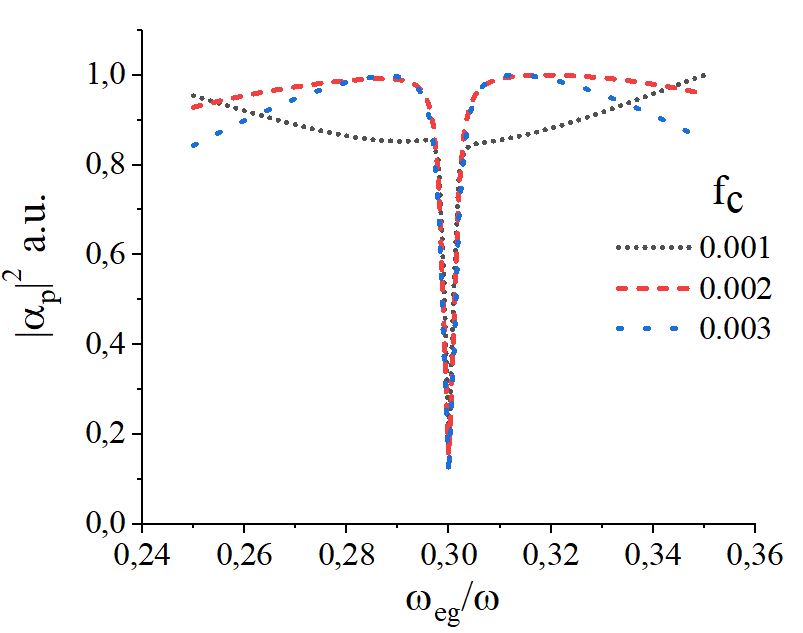}
\caption{MNP plasmon intensity ($|\alpha_p|^2$) versus quantum emitter level-spacing $\omega_{eg}$ calculated for different values of the MNP-QE coupling strength $f_c$. We observe the suppression of the plasmon field at $\omega_{eg}=0.3\omega$. Other parameters are $g=0.01\omega$, $\gamma_p=0.1\omega$,  $\Omega_1=\omega$, $\Omega_2=0.3\omega$, $\Omega_3=0.7\omega$, $\Omega_p= 0.3\omega$, $\gamma_1=\gamma_2=\gamma_3=5\times 10^{-4}~\omega$, $\gamma_{eg}=10^{-5}\omega$,  and $\chi_2=2 \times 10^{-9}~\omega$.  } 
\end{figure} 

In Fig. 5, we plot the MNP plasmon field intensity ($|\alpha_p|^2$) versus level-spacing of the QE $\omega$ for different $f_c$ values. The plasmon intensity demonstrates a dip at $\omega_{eg}$= 0.3$\omega $.


%

When the Figs. 4a and 5 are compared, one can realize that a 300 fold enhancement appears in  the down-converted field $|\alpha_2|^2$, for $\omega_{eg} = 0.3\omega$, when the plasmon mode of the MNP is suppressed. Thus, it is natural to get suspicious if this 300 fold enhancement occurs, actually, due to the suppression of the plasmon excitation around $\omega_{eg}= 0.3\omega$ , i.e., when strong absorption of the MNP is turned off.  We, such a suspicion in mind, checked our calculations several times. We inspected if we define the enhancement factor of $|\alpha_2|^2$, as a mistake, by comparing the two cases, i.e., when (i) MNP-QE dimer is present versus (ii) MNP is present alone. We confirmed that in our all results we compare the case (i) when MNP-QE dimer is present versus (ii) no particle is present (a bare crystal), i.e., the one in Eq.~(11). 

In Fig 6. we change the parameters. After carrying out a neat inspection for the new parameters, we realize that, the maximum of the enhancement does not appear at $\omega_{eg}=0.3\omega$, but it appears at $\simeq 0.297\omega$ in Fig.~6a. In Fig.~6b, however, we observe that plasmon excitation is suppressed maximally still at $\omega_{eg}=0.3\omega$. For other choices of the parameters $f_c$ and $g$, one can see this discrimination more explicitly. That is, maximum $|\alpha_2|^2$ enhancement appears for an $\omega_{eg}$ which is apparently different than the $\omega_{eg}=0.3\omega$ where plasmon absorption is suppressed.

\begin{figure}
 \centering
\includegraphics[width=8cm, height=4.2cm]{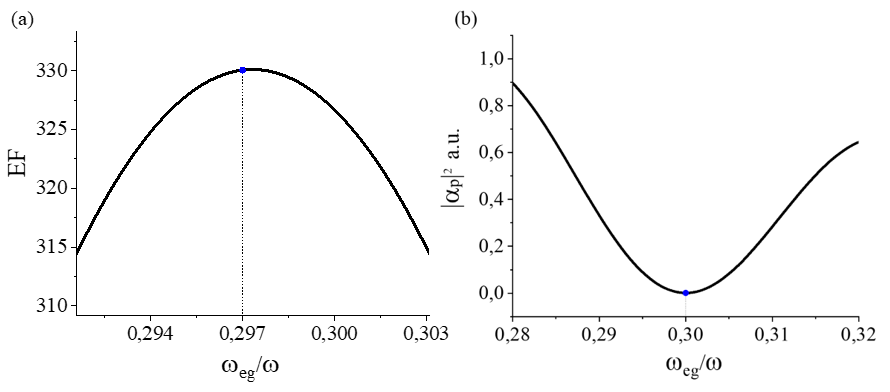}
\caption{ (a) Maximum enhancement factor is observed at $\omega_{eg}\simeq 0.297\omega$ for different parameters. (b) Whereas, the maximum suppression of the MNP excitation is still at $\omega_{eg}=0.3\omega$ for the new parameter set $\gamma_p=0.01~\omega$, $f_c = 0.03\omega$, $g=0.1\omega$, $\Omega_1=\omega$, $\Omega_2=0.3\omega $, $\Omega_3= 0.7\omega$, $\Omega_p= 0.3\omega$, $\gamma_1=\gamma_2=\gamma_3=5\times 10^{-4}\omega$, $\gamma_{eg}=10^{-5}\omega$,  and $\chi_2= $  $2\times 10^{-9}\omega$. } 
\end{figure}

Actually, we face a similar situation in Ref.~\cite{gunay2019fano}, too, where we a study second harmonic converting crystal embedded with MNP-QE dimers. That is, maximum second harmonic generation~(SHG) enhancement appears, again, near the converted frequency ($ \omega_{eg}\simeq 2\omega$) where MNP excitation/absorption is suppressed. This led us a suspicion similar to the one here.  In Ref.~\cite{gunay2019fano}, unlike the present study, we manage to obtain a simple analytical form for the second harmonic field amplitude, which can explain such a behavior. The path interference effects in an unlocalized nonlinear process, e.g. taking place in the body of a crystal, possess an \textit{enriched} cancellation scheme~\cite{gunay2019fano} in the denominator. This is more sophisticated compared to the one where frequency conversion takes place at the hot spots~\cite{turkpence2014engineering}. Eq.~(10), in Ref.~\cite{gunay2019fano}, demonstrates that cancellations in the enriched denominator can take place when the QE level-spacing is near the converted frequency, e.g. $\omega_{eg}\sim 2~\omega$ in Ref.~\cite{gunay2019fano} or $\omega_{eg}\sim 0.3\omega$ here.

It is interesting that the intensity of the other down-converted field, $\alpha_3$, is almost unaffected from the enhancement in the $\alpha_2$ field. That is, the conversion control takes place only in the field, i.e. $\alpha_2$, where the path interference is created. Despite the fact that the hamiltonian and the equations of motion are quite basic/simple, i.e. leaving no doubt on the correctness of our results~\footnote{No approximations are made in the solution of equations of motion. The presented solutions are the time evolution, i.e. exact solutions, of these equations.}, to be honest, the following argument is quite counter-intuitive as well as being intriguing. Yes, the number conservation constraint, originating from the process $\hat{a}_2^\dagger \hat{a}_3^\dagger \hat{a}_1$, between the $|\alpha_2|^2$ and $|\alpha_3|^2$, is broken by the plasmonic (dimer) term in Eq.~(6), and Eqs. (9)-(10). So, one necessarily does not expect an equal intensities for the $\alpha_2$ and $\alpha_3$. 
Still, a counter-intuitiveness is present because: coupling of the $\alpha_2$ mode to a MNP-QD dimer (a system which can  provide merely an absorbing medium) can increase the number of $|\alpha_2|^2$ photons without altering the number of photons in the $|\alpha_3|^2$ down-converted mode.

\section{Summary and Discussions} \label{sec:conclusion}

In this work, we study path interference effects in a nonlinear crystal generating down-converted signals. A long-live quantum emitter, weakly interacting with the down-converted field alone, is made couple with the down-converted field over an MNP’s hot spot. In other words, MNPs are utilized as interaction centers. Analytical results show that it is possible to strengthen the down-converted signal more than 2 orders of magnitude in a nonlinear crystal having an inherently weak output, when the parameters are chosen accordingly. We stress that this enhancement factor further multiplies the enhancement occurring due to the localization field (hot spot) enhancement.

In recent experiments~\cite{plasnap,cuzn,gianten} and the theoretical predictions~\cite {Plasmon_th1,Plasmon_th2}, the pumped cavity mode, $\Omega_1$ mode oscillating with $ e^{-i\omega t}$, is coupled with the MNPs. Field localization effect of MNP is utilized~\cite{plasnap,cuzn,gianten}. In the present study, in difference, the MNPs couple with the down-converted signal $\omega_2 =0.3\omega$. The present work can neither treat the enhancement due to localization, since it does not conduct a simulation of 3D Maxwell equations. The localization effect is referred only as strong light-MNP and MNP-QE couplings.

Retardation effects, in the studied system, can be more demolishing compared to the ones for a localized nonlinear conversion process. Yet, very recently~\cite{gunay2019fano} we managed to demonstrate that retardation effects do not wash out the path interference phenomenon, in a second harmonic converting crystal. Our work presents a first demonstration of the path interference effect in a spontaneous down-converting nonlinear crystal.

While the actual aim of this work is to see the path interference effects, i.e. disassociating the effects of the localization, we can propose following schemes for much stronger spontaneous down conversion enhancements. Using a metal nanostructure, supporting two plasmon modes resonant to $\omega_1 =~\omega$ and $\omega_2=0.3\omega$, could enhance the down conversion process (i) by localizing both fields and (ii) by controlling the enhancement via path interference the QE ($\omega_{eg}$) creates around the lower-energy plasmon mode\footnote{This is a double-resonance scheme \cite{chu2010double}. A triple resonance scheme can be further considered for much stronger enhancement.}, near $\omega_2= 0.3\omega$ . Still, the path interference enhancement multiplies the localization enhancement.

Besides embedding MNP-QE dimers into the crystal, Fig. 1, they can also be coupled to the down-converted cavity mode via evanescent waves, e.g. by positioning the MNP-QE dimers to the maxima lobes, see Fig. 2. This configuration also allows continuous tuning of the QE level spacing, $\omega_{eg}$, via an applied voltage for observing the path interference spectrum, i.e. as in Fig. 2.

There is no doubt that quantum technologies, based on quantum optics/plasmonics phenomena such as entanglement and squeezing, will reshape the new century~\cite{NaturePhot_2019_IntegratedQuantumTechnologies,NaturePhot_2009_QuantumTechnologies,Nanophot_2017_PlasmonicsQuantumTechnologies}. Gaining control over these phenomena or developing an understanding on them (studied in this paper) have crucial importance in the development of new quantum technologies. 
\begin{acknowledgments}
MET and MG acknowledges support from TUBITAK Grant No. 117F118. MET and ZA  acknowledge support from T\"{U}BA-GEB\.{I}P 2017 award.
\end{acknowledgments}

\bibliography{bibliography}

 \end{document}